\documentstyle[12pt]{article}
\newcommand{\bea}{\begin{equation}}
\newcommand{\eea}{\end{equation}}
\newcommand{\ber}{\begin{eqnarray}}
\newcommand{\eer}{\end{eqnarray}}

\begin{document}
\title{Hopf-Turing mixed mode and pattern selection in reaction diffusion systems }
\author{A.Bhattacharyay \\
Department of Theoretical Physics \\
Indian Association for the Cultivation of Science \\
Jadavpur, Calcutta 700 032, India}
\date{\today}
\maketitle
\begin{abstract}
The amplitude equation of Gierer-Mainhardt model has been actually derived near the boundary abuot which Turing and Hopf modes exist. In a parameter region where Hopf-Turing mixed mode solution is stable, a chaotic state that generally results from interaction between mixed modes, is observed. This chaotic region follows a strong selection of a spatially periodic order followed by a local, resonant, very large frequency temporal oscillation. A spatio-temporal forcing, responsible for what obseved, has been identified. 
 \\\\
PACS number(s): 87.10.+e, 47.70.Fw
\end{abstract}
\newpage
Turbulence due to interaction of a spatial (Turing) and temporal (Hopf) modes has caught attention of physicists for more than a decade \cite{Busse,wit}. A Hopf-Turing coupled system shows turbulence of mainly two types, 1) phase turbulene and 2) amplitude turbulence \cite{coullet,kuramoto,brand}. Phase turbulence or Phase chaos, is generally observed in a parameter region where adiabatic elimination of amplitude modes leads to a nonlinear phase equation. This phase turbulence is weak compared to the second type, the amplitude turbulene. The parameter region where amplitude turbulence occurs is that region where adiabatic elimination of amplitude is not possible because phase and amplitude are strongly coupled. A region of defect turbulence \cite{coullet,shraiman}is believed to interpolate between these two regions. Turbulence resulting from interation of mixed modes, or on the other hand stability of different modes against this turbulence in the region where Turing and Hopf modes interact, has been studied on by numerical integration of prototypical amplitude equations. An actual derivation of amplitude equation in the region of interest, as has been done here, not only puts constraints on the variation of relavant coefficients of the equation but also gives the proper selection for the wave number and frequencies over which variations are investigated. Gierer-Meinhardt (GM) model which was originally put forward to explain some observed features of regeneration of hydra \cite{gierer} has a codimension-2 point where two bifurcation boundaries intersect \cite{koch}. One of those boundaries is between a stable spatially periodic (Turing) region and homogeneous osillatory (Hopf) region and is of present interest. Near this boundary a coupled equation for slowly variing Turing and Hopf mode amplitudes has been derived by the use of multiple scale perturbation technique. A numerical investigation of this equation reveals strong selection of a wave number in a region where a spatiotemporal chaos results due to interation of a Turing mode and a travelling wave solution of the coupled amplitude equation. This mode loses stability to a very rapidly growing large frequency local temporal oscillation. This result is particularly interesting because it arises in a region where chaos is believed to prevail and the result is general since it has been obtained from the amplitude equation. In the present latter I want to bring out a forcing, responsible for the resonant selection, that arises from the simultaneous existence of a Turing mode, a travelling wave mode and a homogeneous oscillation in the solution of the above mentioned amplitude equation. 
\par
The Gierer-Meinhardt model in its simplified form \cite{koch,bhatta} is 

\ber\nonumber
\frac{\partial A}{\partial {t}}&=&D{\bigtriangledown} ^2 A +  \frac {A^2}
{B} - A + \sigma   
\eer
\ber
\frac{\partial B}{\partial {t}}&=&{\bigtriangledown} ^2 B +\mu(A^2-B)
\eer 
where A s the activator and B is the inhibitor species, D is the ratio of two diffusion rates, $D_A$ and $D_B$. D is always less than unity since $D_B >> D_A$ which is the Turing condition for fomation of stable pattern \cite{turing}, $\mu$ is ratio of two actual removal rates of A and B respectively, $\sigma$ is the basic production rate of A. From linear stability analysis of Eq.(1) there is a boundary at $\mu = \mu_0$ where $\mu_0 = \frac{1-\sigma}{1+\sigma}$ \cite{bhatta}, above which there is stable a Turing phase and below there is a Hopf phase.
\par
Near the boundary $\mu_0 = \frac{1-\sigma}{1+\sigma}$, a multiple scale perturbation expansion, in two dimensions, of the slow amplitudes of a mixed mode solution of the form $Te^{iK_0x}+He^{i{\omega}_0t}$ results in amplitude equations \cite{cross} os follow \cite{ab}

\ber\nonumber
\frac{\partial T}{\partial {t}}=\epsilon{{\alpha}_r}T+{\beta_r}(2iK_0{{\delta}_x}+{{\delta_y}^2)^2}T+\delta_r{{\left\vert{T}\right\vert}^2}T+\zeta_r{{\left\vert{H}\right\vert}^2}T
\eer
\ber
\frac{\partial H}{\partial {t}}=\epsilon{{\alpha}_c}H+{\gamma_c}{{\delta}_x}^2H+\delta_c{{\left\vert{H}\right\vert}^2}H+\zeta_c{{\left\vert{T}\right\vert}^2}H
\eer
where $\epsilon = \mu-\mu_0 = \mu-\frac{1-\sigma}{1+\sigma}$. The suffix 'r' means coefficients  are real and 'c' indicates that those are complex. All these coefficients are functions of parameters $\sigma$ and D of the GM model. Important point to be noted here is that the equation for the amplitude of Hopf mode is a one dimensional one whereas that for turing mode is two dimensional in space. This is because the Fredholm's condition in the first order of expansion imposes this restriction. The solvability condition also fixes the $K_0$ as $K_0^2 = (\frac{{2D(1-\sigma)}^{1/2}}{1+\sigma}+\frac{1-\sigma}{1+\sigma})/D$. It is evident from Eq.(2) that the interaction between Turing and Hopf modes is essentially one dimensional at least within the scales of derivation of the Eq.(2). In what follows $\omega_0$ has been set equal to $\frac{1-\sigma}{1+\sigma}$ \cite{bhatta} which is its value at the boundary $\mu = \mu_0$ (as given by linear stability analysis).     
\par
Equation (2) has many a solutions, 1) $T = T_0e^{iQ_1x}$ and $H = 0$, 2) $T = 0$ and $H = H_1e^{i\omega_1t}$ 3) $T = 0$ and $H = H_2e^{i(Q_2x-\omega_2t)}$ \cite{perraud} and all sorts of combinations of these modes also exist. We are interested here in the the region where mixed mode solution $T = T_0e^{iQ_1x}$ and $H = H_1e^{i\omega_1t}+H_2e^{i(Q_2x-\omega_2t)}$ is existing. The condition for existence of such a solution is $T_0^2$ positive. $T_0^2$ is positive when $\bigtriangleup > 0 $, where $\bigtriangleup = {\delta_r}{(\delta_c)_{real}}-{\zeta_r}{(\zeta_c)_{real}}$. In such a parameter region where a mixed mode is stable, Band of modes of marginal stability come into play towards generation of deffects what generally turns the system chaotic. Fig(1) shows a plot of $\bigtriangleup$ vs D for $\sigma=0.3$ from which I have picked up the the value $D = 0.04$. Eq.(2) has been numerically integrated by finite difference method at $D=.04$, $\sigma = 0.3$ and $\epsilon = 0.25$ in one dimension on a 64 grid point lattice and on a 128 grid point lattice as well with periodic boundary conditions. Results are shown as space time plots in Fig(2) and Fig(3). Fig(2)a and Fig(3)a show a travelling wave to exist for the first 1000 iterations on both of the lattices. In Fig.(2)b and Fig(3)b (ploted form 1000th to 1600th temporal iterations) we see a particular spatial order to develope, much quickly in time except for a modulation on amplitude in Fig(3)b which is on bigger lattice. The most important point in this spatial structure is that it has the same wave number as the preceding travelling wave. This feature is clarified in Fig(4)a and b where the power spectrum of spatial structures at 1000th and 1600th temporal iterations have been plotted. the power spectrum has been obtained using Microcal Origin4.1 standard FFT tool in Weltch windowing. The biggest peak at about 0.5 in Fig(4)b which corresponds to the rapidly growing spatial order is not only at the same position as the biggest peak in Fig.(4)a corresponding to the travelling wave but also has grown about 100 times. In Fig.(4)b, generation of other modes that renders the system chaotic after some more iterations is also clear. Fig.(4)c shows the homogeneous oscillatory temporal modes, which have been excited during the first 1000 steps of temporal iterations. These small amplitude homogeneous ocsillatory modes are superposed on the travelling waves.  Another point to be noted is that initiation of this particular spatial order happens near the far end of the lattice and about 50 steps of temporal iterations earlier in the longer lattice. This early developement of order in the bigger lattice shows up in relatively bigger heights of the big peaks in Fig.(3)b than in Fig.(2)b. The instability first developes in the far end is evident from the fact that as we go towards the far end of the lattice, the peaks are of bigger size. In Fig.(2)c, a two dimentional concentration map of the smaller lattice, the temporal evolution of the system is from 1625th to 1655 temporal iterations. This figure shows alternate bright and dark spots in two regions near the end which actually represents local temporal oscillations of so much rapid growth that it goes beyond the computation limit after next two temporal iterations.     
\par
In the region where mixed mode solution os Eq.(2) exists, a perturbation of this Eqs.2 as
\ber\nonumber
T = T_0e^{iQ_1x}+\delta{T}
\eer
\ber
H = H_1e^{i\omega_1t}+H_2e^{i(Q_2x-\omega_2t)}+\delta{H}
\eer
results in linear coupled equations in $\delta{T}$ and $\delta{H}$ of the form
\ber
\pmatrix{\delta{T}\cr\delta{H}}=\pmatrix{a&b\cr c&d}\pmatrix{\delta{T}\cr \delta{H}}
\eer
The diagonal terms 'a' and 'd' are of the form $(P_T{T}+Q_T{\delta_x}^2T+R_T{\cos[{Q_2{x}-(\omega_1+\omega_2)t}]})$ and $(P_H{T}+Q_H{\delta_x}^2T+R_H{\cos[{Q_2{x}-(\omega_1+\omega_2)t}]})$ respectively. The off diagonal terms 'b' and 'c' are of the form $(S_T e^{i\phi_1}+U_T e^{i\phi_2})$ and $(S_H e^{-i\phi_1}+U_H e^{-i\phi_2})$ respectively, where P,Q,R,S and U are constants. Off diagonal terms has phases $\phi_1$ and $\phi_2$ which are that of travelling waves and can easily be got rid of by successive transformations of the form
\ber
\pmatrix{\delta T \cr \delta H}=\pmatrix{\delta T \cr e^{-i\phi}\delta H}
\eer
The $\cos[Q_2x-(\omega_1+\omega_2)]$ forcing in the coefficients of the diagonal terms  make each part of Eq.(4) a Mathue-equation \cite{jordan} in terms of spatial co-ordinates. Interestingly this force is of wave number same as that of the travelling wave which is seen to be the fact in the results of the numerical integration. As a consiquence of this spatial Mathue a steady pattern developes at the forcing wave number. Spatial pattern results more prominently at furthest end since it should grow exponentially faster at large distance according to the nature of standard unbounded solution of Mathew equation. The pattwern first developes at the farthest end and afterwards it grows rapidly in the whole lattice due to periodic boundary conditations. The early growth of this typical spatial order in larger lattice is a consiquence of its very origin. If one carefully looks at into the shape of Eq.4, it is evident that there is a temporal forcing resulting in a Mathue-equation, but in the next higher order derivatives with respect to time. To my opinion this temporal Mathue is the cause of late developement of local, resonant, temporal oscillations as shown in Fig.(2)c . This typical selection mechanism has its root in simultaneous existence of a travelling wave and homogeneous oscillation along with a turing mode as a  solution of Eq.(2). Only near this type of stable solution of Eq.(2) if it is linearised the nonlinear terms results in the forcing terms as mentioned above.
\par
Onset of spatio-temporal chaos as a result of interaction of Turing and Hopf modes, near a boundary between these two phases, is presently coming forth as an important scenario in the context of chaos in reaction-diffusion systems. Existence of a mixed mode comprising of a Turing and a travelling wave suffices for onset of chaos. A homogeneous oscillation will  always exist along with the above mentioned mixed mode when the coefficient of the linear term in the Hopf part of Eq.(2) is complex with positive real part. A real coefficient of that linear term will not in general make the homogeneous oscillation exist in the mixed mode solution of Eq.(2). The point of interest of present letter is that the existence of a homogeneous oscillation along with above mentioned mixed modes results in a forcing, spatially as well as temporally, and that plays for selection of spatial and temporal orders in the region where chaos should prevail. Besides all these, since there are Mathue equations, there always remains a chance of parametric resonance at double the wave length and time period as that of forcing. Controlling chaos \cite{ott1} by applycation of small perturbations is a much studied topic these days. Existence of strange attractor enhances the possibility of fixing varied states without costly alterations of a given model. Strange attractors, being dense set of unstable peiodic orbits \cite{ott2}, are destabilised by resonating one of those unstable oscillations with small external stimulants to bring out an order from chaos. So, when a system does have a Mathue equation in it, in a region where the system is turbulent, an adjustment in the forcing frequencies can practically pick up an order from chaos with small perturbations. Thus the characteristic feature of the present system has its importance in its possibility of getting tuned to prefer an order to chaos. This very characteristic feature of the reaction-diffusion systems also establishes its robastness against chaos.
\section*{Acknowledgment}
I acknowledge very useful discussions and help that I have got from my supervisor Prof. J.K.Bhattacharjee and I also acknowledge the help I have received from my friend Kausik Sankar Das in learning some computational details etc.
\newpage\section*{Figure Caption}
{\bf Fig.(1)} is a plot of $\bigtriangleup vs D$ for $\sigma = 0.3$, from which the region of interest has been chosen at $D=0.04$.
\newline
{\bf Fig.(2)a} is an amplitude vs time plot of 'H' (Hopf mode) for first 1000 temporal iterations of the 64 point space lattice where relatively brighter regions are regions of greater concentrations.
\newline
{\bf Fig.(3)a} is an amplitude vs time plot of 'H' (Hopf mode) for first 1000 temporal iterations of the 128 point space lattice where relatively brighter regions are regions of greater concentrations.
\newline
{\bf Fig.(2)b} is a 3-dimensional amplitude vs time plot of 'H' (Hopf mode) form 1001 to 1600th  temporal iterations of the 64 point space lattice where prominant growth of a spatial order is present.
\newline  
{\bf Fig.(3)b} is a 3-dimensional amplitude vs time plot of 'H' (Hopf mode) form 1001 to 1600th  temporal iterations of the 128 point space lattice where prominant growth of a spatial order, except for an amplitude modulation, is also present.
\newline
{\bf Fig.(2)c} is an amplitude vs time plot of 'H' (Hopf mode) for 1625th to 1655th temporal iterations of the 64 point space lattice. In this Figure alternate bright and dark spots along the time axis near its end are representatives of large amplitude large frequency temporal oscillations.
\newline
{\bf Fig.(4)a} is a power spectrum for the spatial orders at the end of 1000th temporal iteration of the 64 point space lattice. 
\newline
{\bf Fig.(4)b} is a power spectrum for the spatial orders at the end of 1600th temporal iteration of the 64 point space lattice.              
\newline
{\bf Fig.(4)c} is an FFT to show the homogeneous temporal orders at the end of 1000th temporal iteration of the 64 point space lattice.

\enddocument
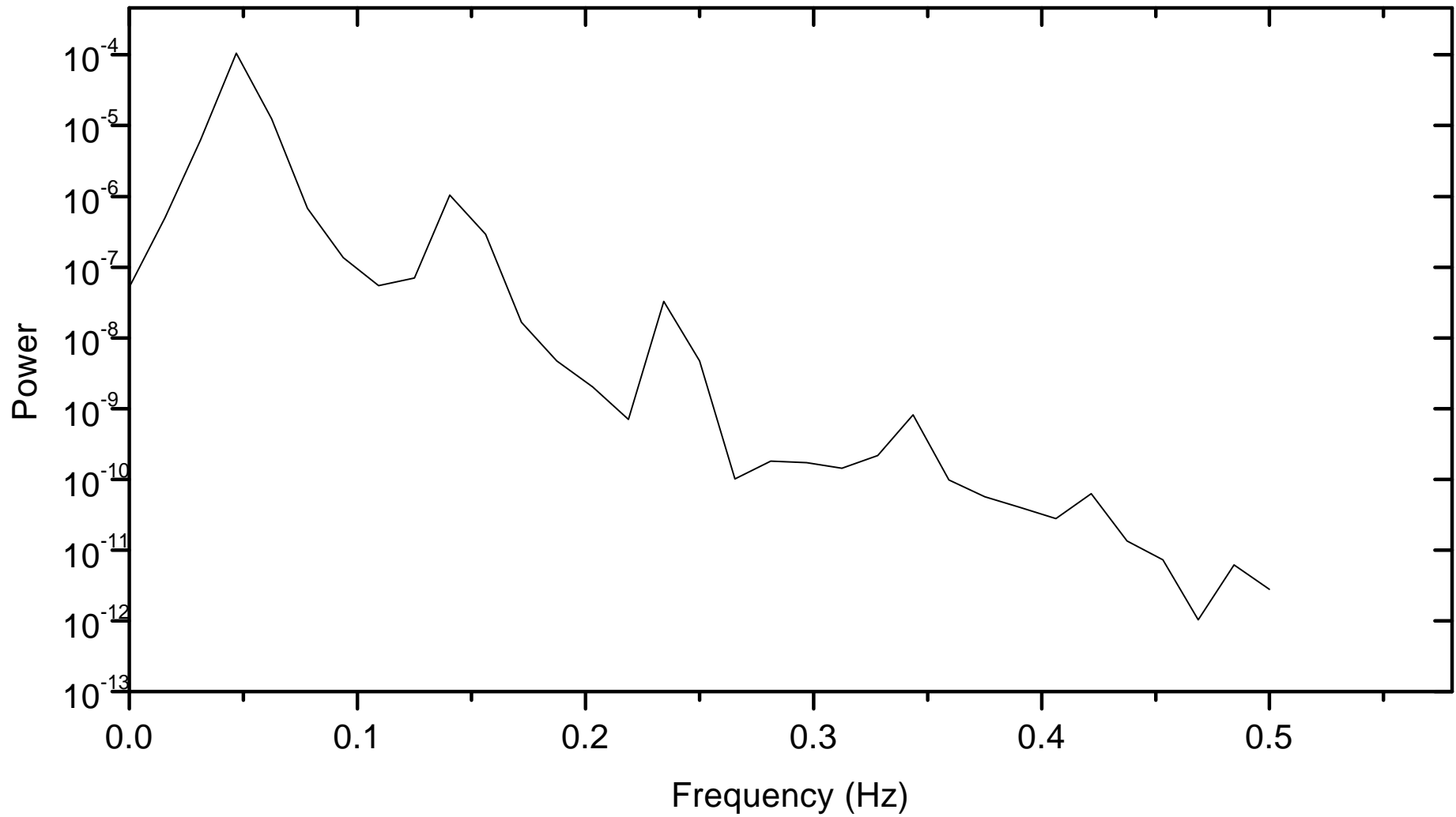
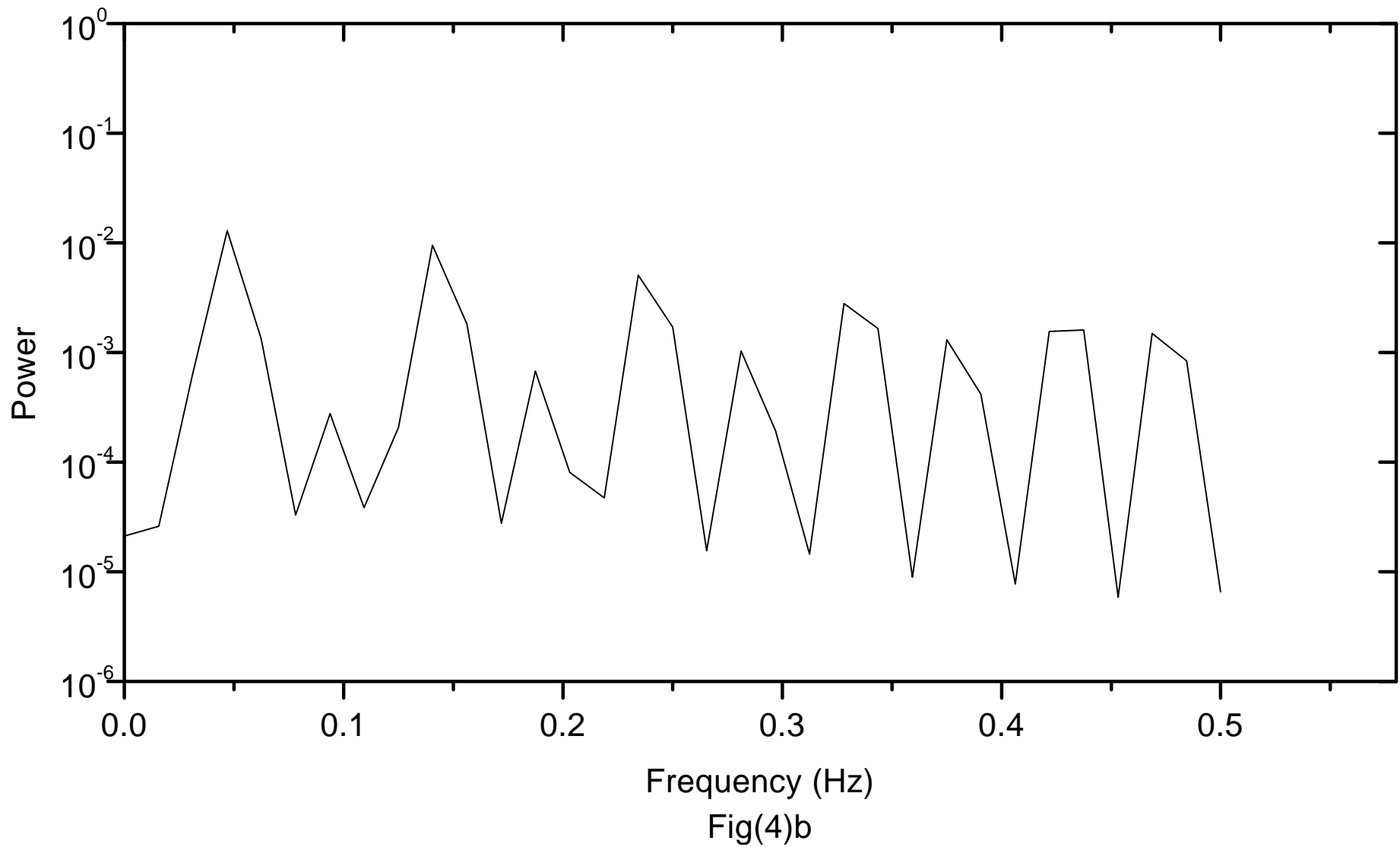
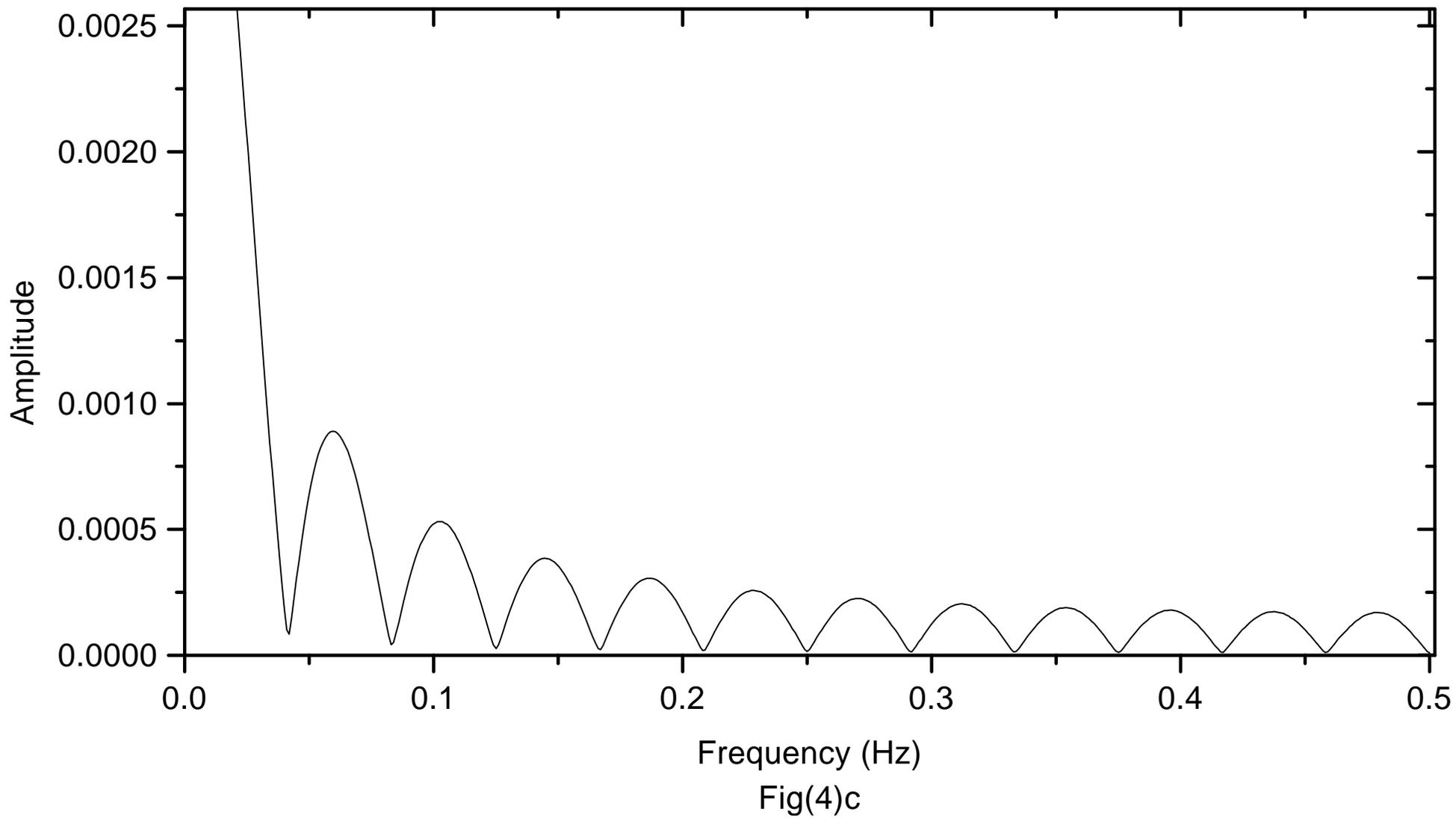
